\documentclass[aps,prb,twocolumn,preprintnumbers,amsmath,amssymb]{revtex4-1}
\usepackage{graphicx}
\usepackage{dcolumn}
\usepackage{bm}
\usepackage{amsmath}
\usepackage{color}

\begin{document}

\def\gsim{\mathop {\vtop {\ialign {##\crcr 
$\hfil \displaystyle {>}\hfil $\crcr \noalign {\kern1pt \nointerlineskip } 
$\,\sim$ \crcr \noalign {\kern1pt}}}}\limits}
\def\lsim{\mathop {\vtop {\ialign {##\crcr 
$\hfil \displaystyle {<}\hfil $\crcr \noalign {\kern1pt \nointerlineskip } 
$\,\,\sim$ \crcr \noalign {\kern1pt}}}}\limits}


\title{
%
Dynamical and Static Structure Factors in Hedgehog-Antihedgehog Order in Icosahedral 1/1 Approximant Crystal}


\author{Shinji Watanabe$^{*}$}
\affiliation{Department of Basic Sciences, Kyushu Institute of Technology, Kitakyushu, Fukuoka 804-8550, Japan 
}


\date{\today}

\begin{abstract} 
Recent discoveries of magnetic long-range orders in the icosahedral quasicrystal and topological magnetic structures on the icosahedron (IC) as the hedgehog state and the antihedgehog state have attracted great interest. 
Here, we report our theoretical analysis of the dynamical as well as static structure of the hedgehog-antihedgehog order in the 1/1 approximant crystal (AC).
By constructing the effective magnetic model for the rare-earth based AC, 
on the basis of the linear spin-wave theory, the excitation energy is shown to exhibit the reciprocal dispersion, as a consequence of preservation of the spatial inversion symmetry by the hedgehog-antihedgehog ordering. 
The static structure factor is shown to be expressed generally in the convolution form of the lattice structure factor and the magnetic structure factor on the IC(s) and the numerical calculation reveals the extinction rule.  
The dynamical structure factor shows that the high intensities appear in the low-energy branch along the $\Gamma$-X line and the R-$\Gamma$-M line in the reciprocal space.  
\end{abstract}


\maketitle

\section{Introduction}

Quasicrystal (QC) has no periodicity of the lattice but possesses unique rotation symmetry forbidden in periodic crystals. 
Since the discovery of the QC~\cite{Shechtman}, the understanding of the lattice structure has proceeded~\cite{Tsai,Takakura}. 
However, the electronic state in the QC is far from complete understanding because the Bloch theorem based on the translational invariance can no longer be applied to the QC. 
Hence, the clarification of the electronic states and the physical property in the QC is the frontier of the condensed matter physics. 

One of the interesting questions about the QC has been whether the magnetic long-range order is realized in the three-dimensional QC. 
Experimentally, the magnetic long-range order has been explored in the QC and also in the approximant crystal (AC) whose lattice is composed of the common local atomic structure to that in the QC with periodicity.
In the rare-earth based 1/1 AC where the 4f electrons are responsible for the magnetism, the magnetic long-range order has been observed by the bulk measurements such as the magnetic susceptibility and the specific heat ~\cite{Suzuki2021}.  
The antiferromagnetic (AFM) order has been observed in the 1/1 AC Cd$_6$R (R=Tb, Y, Pr, Nd, Sm, Gd, Tb, Dy, Ho, Er, Tm, Yb, and Lu) ~\cite{Tamura2010,Mori} and in the 1/1 AC Au-Al-R (R=Gd and Tb)~\cite{Ishikawa2018}. 
The ferromagnetic (FM) order has been observed 
in the 1/1 AC Au-SM-R (SM=Si, Ge, and Al; R=Gd, Tb, Dy, and Ho)~\cite{Hiroto2013,Hiroto2014,Ishikawa2016}
{(Note that Au-SM-R is the composition tunable compound which exhibits the FM order and AFM order depending on the composition ratio of Au and SM, e.g., Au$_x$Al$_{86-x}$Gd~[10].)}.
As for the QC, the spin-glass behavior has been observed in the QC Cd-R (R=Gd, Tb, Dy, Ho, Er, and Tm)~\cite{Goldman}. 
Recently, the FM orders have been discovered in the QC Au$_{65}$Ga$_{20}$R$_{15}$ (R=Gd and Tb)~\cite{Tamura2021} and in Au$_x$Ga$_{85-x}$Dy$_{15}$ $(x=62$-$68)$~\cite{Takeuchi2023}, which have brought about the breakthrough. 

Among these ACs and QCs, the neutron measurements have been performed recently in some Tb-based ACs and Ho-based AC, which have revealed the non-collinear and non-coplanar alignments of the 4f magnetic moments 
of the ferrimagnet in the 1/1 ACs Au$_{70}$Si$_{17}$Tb$_{13}$~\cite{Hiroto} and Au-Si-R (R=Tb, Ho)~\cite{Gebresenbut} as well as the antiferromagnet in the 1/1 AC Au$_{72}$Al$_{14}$Tb$_{14}$~\cite{Sato2019}.

\begin{figure}[t]
\includegraphics[width=7cm]{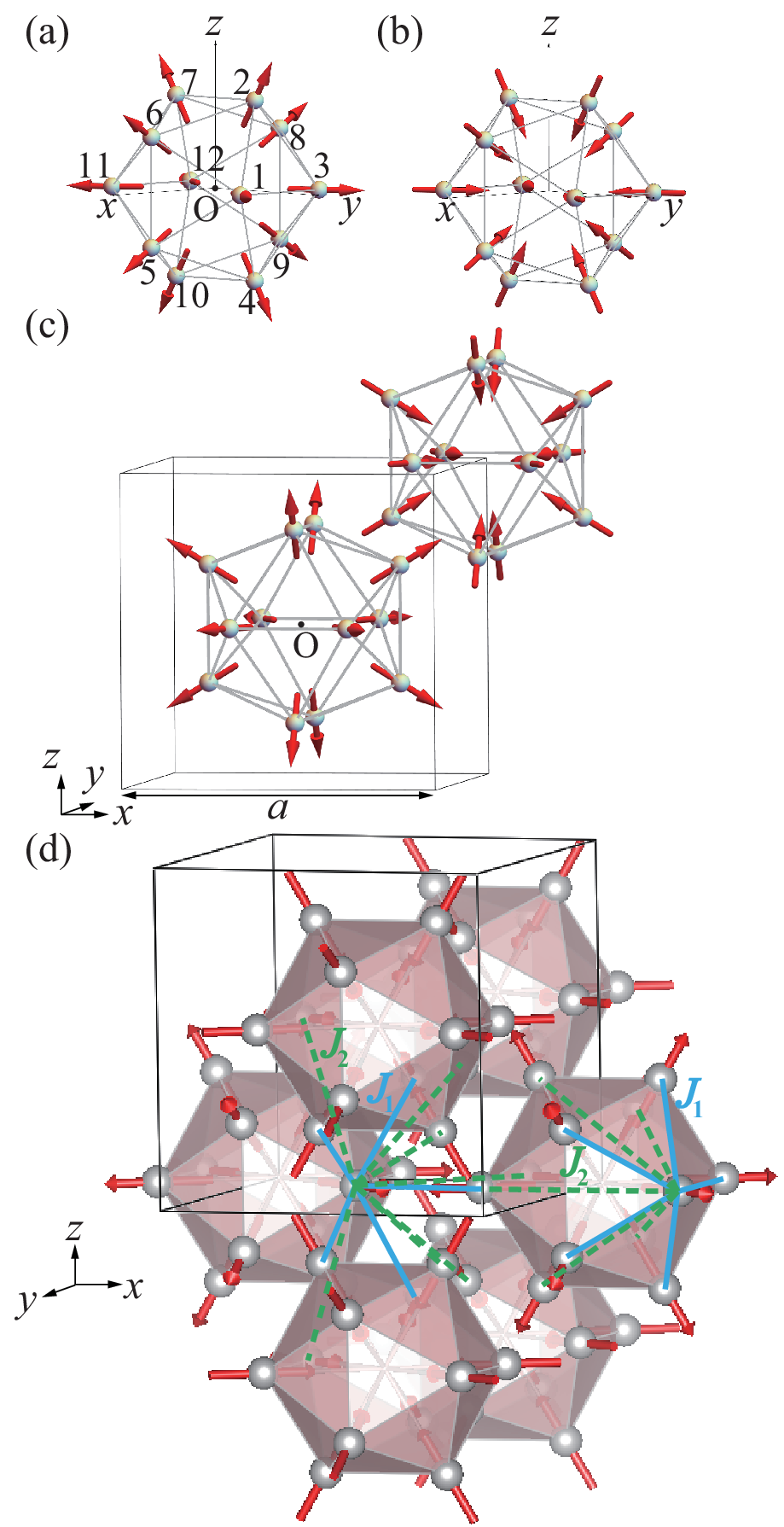}%
\caption{
(a) The hedgehog state on the IC. The number labels the Tb sites at the 12 vertices of the IC.  
(b) The antihedgehog state on the IC.
(c) The hedgehog-antihedgehog order in the 1/1 AC. 
The frame box is the bcc unit cell.  
{
(d) The nearest-neighbor (N.N.) interactions $J_1$ (blue line) and next-nearest-neighbor (N.N.N.) interactions $J_2$ (green dashed line) for the intra IC and the inter IC are illustrated. 
}
}
\label{fig:HAH}
\end{figure}

These ACs and the QCs are composed of the concentric shell structures of atomic polyhedrons, which is referred to as the Tsai-type cluster~\cite{Tsai,Takakura}. 
Inside the Tsai-type cluster, the rare-earth atoms are located at the 12 vertices of the icosahedron (IC) [see Fig.~\ref{fig:HAH}(a)]. 
To understand the magnetic property of the rare-earth based icosahderal QCs and ACs, the clarification of the crystalline electric field (CEF) at the rare-earth site is necessary. 

Recently, theoretical formulation of the CEF at the rare-earth site in the QC and AC has been developed on the basis of the point charge model~\cite{WK2021}. 
By applying this formulation to the QC Au-Al-Yb~\cite{WK2021} and Au-SM-Tb~\cite{WSR,WPNAS}, the CEF was analyzed theoretically. Then, it was shown that 
the CEF ground state possesses the uniaxial anisotropy at each rare-earth site, giving rise to the unique magnetic structures on the IC~\cite{WSR,WPNAS}. 
Interestingly, the magnetic structures which have finite topological charge $n$ defined on the IC such as the hedgehog state displayed in Fig.~\ref{fig:HAH}(a) have been shown to appear. 
The topological charge has the physical meaning of the number of covering the whole sphere by the magnetic moments on the IC. 
The hedgehog state, where all 12 magnetic moments are directed outward from the IC shown in Fig.~\ref{fig:HAH}(a),  has $n=+1$~\cite{WSR}.
The antihedgehog state, where all 12 magnetic moments are directed inward from the IC as displayed in Fig.~\ref{fig:HAH}(b), has $n=-1$. 
The hedgehog state is regarded as the source of the emergent field while the antihedgehog state is regarded as the sink of the emergent field~\cite{Kanazawa,Tokura2021}. 

Then, the effective model for the magnetism in the 1/1 AC and QC, where the magnetic easy axis at each Tb site arising from the CEF is taken into account, was constructed~\cite{WSR,WPNAS}. 
By numerical calculations in the effective model applied to the Cd$_{5.7}$Yb-type QC~\cite{Takakura}, it has been shown that the uniform hedgehog state is stabilized in the ground-state phase diagram~\cite{WSR}.  
This is the theoretical discovery of the topological magnetic state in the QC, which has attracted great interest. 
Moreover, the recent analysis of the dynamical structure factor in the QC has revealed that the non-reciprocal excitation in the uniform hedgehog order emerges in the vast extent of the wave vector ${\bm q}$-energy $\omega$ plane, i.e., $S({\bm q},\omega)\ne S(-{\bm q},\omega)$~\cite{WHH2022}.   

To get insight into the emergence mechanism of the nonreciprocal excitation, the dynamical structure factor in the 1/1 AC has been analyzed recently by assuming the uniform hedgehog order~\cite{W2023}.  
Then, 
the non-reciprocal energy dispersion of magnon has been shown to appear as a consequence of the inversion symmetry breaking by the uniform hedgehog ordering~\cite{W2023}.  

In the ground-state phase diagram of the effective model applied to the 1/1 AC, it was shown that the hedgehog state forms the AFM order~\cite{WSR,WPNAS}. 
Namely, the hedgehog state is formed on the IC located at the center of the unit cell of the body-centered cubic (bcc) lattice while the antihedgehog state is formed on the IC located at the corner of the bcc unit cell as shown in Fig.~\ref{fig:HAH}(c)~\cite{WSR,WPNAS}.  
In this study, we theoretically analyze the dynamical as well as static structure factor in the AFM hedgehog order in the 1/1 AC as the true ground state. 
It is shown that the reciprocal excitation emerges, which is in sharp contrast to the case of the uniform hedgehog order in the 1/1 AC.  
This is understood as the alternative distribution of the hedgehog state and antihedgehog state preserves the inversion symmetry in the 1/1 AC.

Recently, the dynamical structure factors for the topological magnetic textures such as the Skyrmion and the hedgehog in periodic crystals have been studied~\cite{Kato2021,Eto2022}, which have attracted interest. 
As for the QC, the lattice dynamics was studied in the Zn-Mg-Sc icosahedral QC and the 1/1 AC by inelastic X-ray- and neutron-scattering experiments~\cite{Boissieu}. 
Theoretically, the dynamical spin structure factors were studied in the low-dimensional systems such as the Fibonacci chain~\cite{Ashraff1989} and the two dimensional QCs~\cite{Ashraff1990, Wessel2005}.  
Hence, the present study in the three dimensional icosahedral AC considering the magnetic anisotropy arising from the CEF is expected not only to contribute to the understanding the dynamics of the topological magnetic texture but also to get insight into the unique magnetic structure in the icosahedral QC.

The organization of this paper is as follows: In section~II, we introduce the effective model for magnetism in the rare-earth based 1/1 AC. In section~III, we analyze the static structure factor of magnetism in the hedgehog-antihedgehog order in the 1/1 AC. In section~IV, we explain the linear spin-wave theory applied to the 1/1 AC and show the results of the excitation energy and the dynamical structure factor in the hedgehog-antihedgehog order in the 1/1 AC. In section~V, we summarize the paper and discuss the future issues. 


\section{Effective model for magnetism}

By taking into account the effect of the magnetic anisotropy arising from the CEF, 
the effective minimal model for magnetism in the rare-earth based QC and AC has been constructed~\cite{WSR,WPNAS}.  
In this study, we consider the effective minimal model 
\begin{eqnarray}
H=\sum_{\langle i,j\rangle}J_{ij}{\bm S}_i\cdot{\bm S}_j-D\sum_{i}({\bm S}_i\cdot\hat{\bm e}^{i}_{3})^2
\label{eq:H}
\end{eqnarray}
in the 1/1 AC.  As the atomic coordinates, we employ the lattice structure of the 1/1 AC identified 
{by the X-ray measurement}
in Au$_{70}$Si$_{17}$Tb$_{13}$~\cite{Hiroto}. 
The frame box shown in Fig.~\ref{fig:HAH}(c) represents the unit cell of the bcc lattice with the lattice constant $a=14.726~{\rm \AA}$ where the two ICs are located at the center and the corner.

In Eq.~(\ref{eq:H}), 
${\bm S}_i$ represents the magnetic moment of the total angular momentum $J=6$ at the $i$th Tb site with $S_i=6$, which is referred to as ``spin'' hereafter. 
The second term in Eq.~(\ref{eq:H}) expresses the effect of the uniaxial anisotropy due to the CEF, where $\hat{\bm e}^{i}_{3}$ is the unit vector directed to the magnetic easy axis at each Tb site (see section IV and also Fig.~\ref{fig:local_e123} below).

In Eq.~(\ref{eq:H}), $J_{ij}$ is set to be the nearest neighbor (N.N.) interaction $J_1$ and the next N.N. (N.N.N.) interaction $J_2$ for the intra IC and also for the inter IC. 
{
In ref.~\cite{WPNAS}, the ground-state phase diagram of the model (\ref{eq:H}) for the large-$D$ limit was discussed, where the stability of the ground state in the 1/1 AC was analyzed by comparing the energy of the uniform distribution of the magnetic ground state on the IC and the staggered distribution. Recently, the numerical calculation for the ground state of the model for the large-$D$ limit in the 1/1 AC has been performed without assuming the ground state on the IC~\cite{W2023b}. Namely,
i}n 
the large-$D$ limit of the model (\ref{eq:H}), the ground state phase diagram was determined by numerical calculations, where the AFM hedgehog order, i.e., the hedgehog-antihedgehog order shown in Fig.~\ref{fig:HAH}(c) is realized for $J_2/J_1< 0.284$ with the FM interaction $J_1<0$~\cite{WPNAS,W2023b}. 
The strong N.N. interaction $J_{1}$ stabilizes the hedgehog state and antihedgehog state at the center IC and corner IC in the bcc unit cell in Fig.~\ref{fig:HAH}(c) not only for the intra IC but also for the inter IC. 
In this paper, we take $J_1$ as the energy unit, i.e., set $J_1=-1.0$ and mainly show the results for $J_2=-0.1$ as the typical parameter.  

We note that the effective model which corresponds to the large-$D$ limit of the model (\ref{eq:H}) succeeded in explaining the magnetic structures of the FM order of the ferrimagnetism on the IC observed in the 1/1 AC Au$_{70}$Si$_{17}$Tb$_{13}$~\cite{Hiroto} and the AFM order of the whirling-antiwhirling states on the ICs observed in the 1/1 AC Au$_{72}$Al$_{14}$Tb$_{14}$~\cite{Sato2019,WPNAS}. 
The model (\ref{eq:H}) is expected to be relevant to not only the Tb-based AC but also the broad range of the rare-earth based-ACs. 

The interaction $J_{ij}$ in the first term of Eq.~(\ref{eq:H}) set in this study is explicitly noted as follows: 
$J_1$ is set for 5 bonds for each Tb site with the bond length $0.374a$ (1 bond) and $0.378a$ (4 bonds) and $J_2$ is set for 5 bonds for each Tb site with the bond length $0.610a$ (4 bonds) and $0.612a$ (1 bond) for the intra IC 
[see Fig.~\ref{fig:HAH}(d)].
For the inter IC, $J_1$ is set for 5 bonds for each Tb site with the bond length $0.368a$ (4 bonds) and $0.388a$ (1 bond) and $J_2$ is set for 7 bonds with the bond length $0.528a$ (2 bonds), $0.530a$ (4 bonds), and $0.539a$ (1 bond) 
[see Fig.~\ref{fig:HAH}(d)].

{
In this paper, we analyze the property of the dynamics of the hedgehog-antihedgehog ordered state shown in Fig.~\ref{fig:HAH}(c) by applying the liner spin-wave theory to the model (\ref{eq:H}) for large $D$. Namely, the excitation from the ground state of the classical limit of the model (\ref{eq:H}) is analyzed. 
}
Hence, $\hat{\bm e}_{3}^i$ in the second term of Eq.~(\ref{eq:H}) is set as $(\pm\frac{1}{\sqrt{\tau+2}},\pm\frac{\tau}{\sqrt{\tau+2}},0)$ for the $i=1, 3, 11$, and 12th site on the IC in Fig.~\ref{fig:HAH}(a), 
$(\pm\frac{\tau}{\sqrt{\tau+2}},0,\pm\frac{1}{\sqrt{\tau+2}})$ for the $i=5, 6, 8$, and 9th site, and $(0,\pm\frac{1}{\sqrt{\tau+2}},\pm\frac{\tau}{\sqrt{\tau+2}})$ for $i=2, 4, 7$, and 10th site, respectively, 
where $\tau$ is the golden mean $\tau=(1+\sqrt{5})/2$.

\section{Static structure factor of magnetism}

First, let us analyze the static structure factor 
\begin{eqnarray}
F_{s}({\bm q})=\left\langle\left|
\frac{1}{N}\sum_{i}{\bm S}_i e^{i{\bm q}\cdot{\bm r}_i}
\right|^2   
\right\rangle, 
\label{eq:Fs}
\end{eqnarray}
for the ground state of the hedgehog-antihedgehog order in the 1/1 AC.  
Here, $N$ is the total number of the Tb sites and ${\bm q}$ is the wave vector.  
Since the position vector of the $i$th Tb site ${\bm r}_i$ is expressed as ${\bm r}_i={\bm R}_j+{\bm r}_{0m}$ where ${\bm R}_j$ is the position vector of the center of the $j$th unit cell [see Fig.~\ref{fig:HAH}(c)] and ${\bm r}_{0m}$ is the position vector of the $m$th Tb site inside the unit cell $(m=1, 2, \cdots, 24)$, Eq.~(\ref{eq:Fs}) is expressed as the convolution form 
\begin{eqnarray}
F_{s}({\bm q})=F_{L}({\bm q})S_{2{\rm IC}}({\bm q}).
\label{eq:FLS}
\end{eqnarray}
Here, $F_{L}({\bm q})$ is the structure factor of the lattice and $S_{2{\rm IC}}({\bm q})$ is the magnetic structure factor of the 2 ICs inside the unit cell, which are defined as
\begin{eqnarray}
F_{L}({\bm q})&=&\frac{1}{N_{L}^2}\sum_{j=1}^{N_L}\sum_{j'=1}^{N_L}e^{i{\bm q}\cdot({\bm R}_j-{\bm R}_{j'})}, 
\label{eq:FL}
\\
S_{2{\rm IC}}({\bm q})&=&\frac{1}{24^2}\sum_{m=1}^{24}\sum_{m'=1}^{24}\langle{\bm S}_{m}\cdot{\bm S}_{m'}\rangle e^{i{\bm q}\cdot({\bm r}_{0m}-{\bm r}_{0m'})},
\label{eq:S2IC}
\end{eqnarray}
respectively. Here, $N_{L}=N_{1}^3$ is the number of the unit cell and $N=24N_{L}$ holds. 

The position vector ${\bm R}_j$ is expressed as
\begin{eqnarray}
{\bm R}_j=n_1{\bm a}_1+n_2{\bm a}_2+n_3{\bm a}_3, 
\label{eq:R}
\end{eqnarray}
where  ${\bm a}_i$ is the basic translation vector of the lattice defined by ${\bm a}_1=a(1,0,0)$, ${\bm a}_2=a(0,1,0)$, and ${\bm a}_3=a(0,0,1)$ with integer $n_i$ $(i=1, 2, 3)$.
The wave vector ${\bm q}$ is expressed as
\begin{eqnarray}
{\bm q}=h{\bm b}_1+k{\bm b}_2+l{\bm b}_3,
\label{eq:q}
\end{eqnarray}
where ${\bm b}_i$ is the basic translation vector of the reciprocal lattice defined by ${\bm b}_1=\frac{2\pi}{a}(1,0,0)$,  ${\bm b}_2=\frac{2\pi}{a}(0,1,0)$, and  ${\bm b}_3=\frac{2\pi}{a}(0,0,1)$. 
Under the periodic boundary condition along the $N_1{\bm a}_i$ $(i=1, 2, 3)$ directions, $h$, $k$, and $l$ are given by $h=\frac{m_1}{N_1}$, $k=\frac{m_2}{N_1}$, and $l=\frac{m_3}{N_1}$, respectively with integer $m_i$ $(i=1, 2, 3)$. 

By substituting Eqs.~(\ref{eq:R}) and (\ref{eq:q}) into Eq.~(\ref{eq:FL}), $F_L({\bm q})$ is calculated as 
\begin{eqnarray}
F_{L}({\bm q})=\frac{1}{N_L^2}\frac{\sin^2(\pi hN_1)}{\sin^2(\pi h)}\frac{\sin^2(\pi kN_1)}{\sin^2(\pi k)}\frac{\sin^2(\pi lN_1)}{\sin^2(\pi l)}. 
\label{eq:FL2}
\end{eqnarray}
For integer $h$, $k$, and $l$, Eq.~(\ref{eq:q}) becomes the reciprocal lattice vector ${\bm q}={\bm Q}_{hkl}\equiv h{\bm b}_1+k{\bm b}_2+l{\bm b}_3$.   
Then, Eq.~(\ref{eq:FL2}) leads to $F_L({\bm Q}_{hkl})=1$. 
As $h$, $k$, and $l$ deviate from the integer values, $F_L({\bm q})$ decays rapidly in the finite-size system. 
In the bulk limit, namely for $N_1\to\infty$, we obtain 
\begin{eqnarray}
F_L(\bm q)=\delta_{{\bm q},{\bm Q}_{hkl}}.
\label{eq:FL_bulk}
\end{eqnarray}
This implies that from the convolution form of Eq.~(\ref{eq:FLS}), the static structure factor $F_s({\bm q})$ can have the non-zero value for integer $h$, $k$, and $l$ where $F_{L}({\bm q})$ becomes non zero. 

\begin{figure}[t]
\includegraphics[width=7cm]{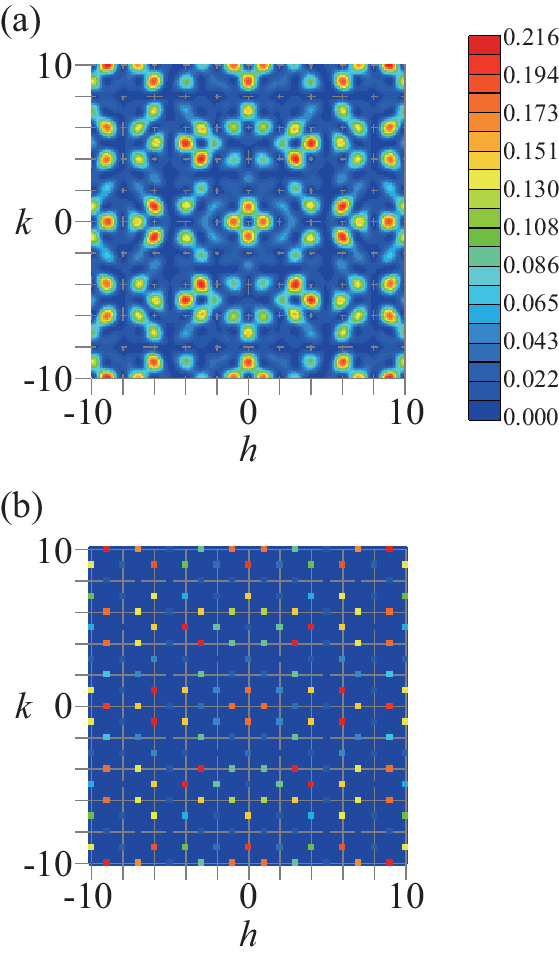}%
\caption{
(a) The intensity plot of the magnetic structure factor for the 2 ICs $S_{2{\rm IC}}({\bm q})$ for ${\bm q}=\frac{2\pi}{a}(h,k,0)$ in the $h$-$k$ plane. 
(b) The intensity plot of the magnetic structure factor $F_s({\bm q})$ for ${\bm q}=\frac{2\pi}{a}(h,k,0)$ in the $h$-$k$ plane. 
The color scale is common to that in (a). 
}
\label{fig:Sq}
\end{figure}

Next, let us analyze the magnetic structure factor of the 2 ICs inside the unit cell [see Fig.~\ref{fig:HAH}(c)] $S_{2{\rm IC}}({\bm q})$ defined in Eq.~(\ref{eq:S2IC}). 
We have calculated $S_{2{\rm IC}}({\bm q})$ in Eq.~(\ref{eq:S2IC}) in the system for $N_L=N_1^3$ with $N_1=256$ numerically. 
By searching the maximum of $S_{2{\rm IC}}({\bm q})$ for ${\bm q}=h{\bm b}_1+k{\bm b}_2+l{\bm b}_3$ with integers $h, k, l\in[-10, 10]$ numerically, we find that the maximum value $0.2158$ appears at the 12 ${\bm q}$ points with $(h,k,l)=(\pm10, 0, \pm 9)$, $(0, \pm 9, \pm 10)$, and $(\pm 9, \pm 10, 0)$.  
We plot  $S_{2{\rm IC}}({\bm q})$ for ${\bm q}=\frac{2\pi}{a}(h,k,0)$ in the $h$-$k$
plane in Fig.~\ref{fig:Sq}(a). 
We have confirmed that $S_{2{\rm IC}}({\bm q})=0$ for ${\bm q}=\frac{2\pi}{a}(h,k,l)$ 
in case that all $h$, $k$, and $l$ are even integers or one of $h$, $k$, and $l$ is even integer.
We have also calculated $S_{2{\rm IC}}({\bm q})$ for ${\bm q}=\frac{2\pi}{a}(0,k,l)$ and for  ${\bm q}=\frac{2\pi}{a}(h,0,l)$. The results are the same as Fig.~\ref{fig:Sq}(a) with $h$ and $k$ being replaced with $k$ and $l$ respectively and with $h$ and $k$ being replaced with $l$ and $h$ respectively. 
The systematic absence of reflection $S_{2{\rm IC}}({\bm q})$ occurs for ${\bm q}=\frac{2\pi}{a}(h,k,l)$ 
for all or one of even integer among $h$, $k$, and $l$. 

From Eq.~(\ref{eq:FLS}) and Eq.~(\ref{eq:FL_bulk}), it turns out that $F_s({\bm q})$ in the bulk limit is equivalent to $S_{2{\rm IC}}({\bm q})$ for ${\bm q}=h{\bm b}_1+k{\bm b}_2+l{\bm b}_3$ with integers $h$, $k$, and $l$. 
Hence, the maximum of $F_s({\bm q})$ appears at $(h,k,l)=(\pm10, 0, \pm 9)$, $(0, \pm 9, \pm 10)$, and $(\pm 9, \pm 10, 0)$ for $h$, $k$, $l\in[-10,10]$.  
We plot the static magnetic structure factor $F_s({\bm q})$ for ${\bm q}=\frac{2\pi}{a}(h,k,0)$ in the 
$h$-
$k$
plane in Fig.~\ref{fig:Sq}(b). 
As noted above,  $F_s({\bm q})$ for ${\bm q}=\frac{2\pi}{a}(0,k,l)$ and  ${\bm q}=\frac{2\pi}{a}(h,0,l)$ are obtained by replacing $h$ and $k$ in Fig.~\ref{fig:Sq}(b) with $k$ and $l$ respectively and with 
$l$ 
and 
$h$ 
respectively.
The systematic absence of reflection $F_s({\bm q})=0$ occurs for ${\bm q}=\frac{2\pi}{a}(h,k,l)$ for the even integers $h$, $k$, and $l$ and also for the case that the two of $h$, $k$, and $l$ are 
odd integers.

\section{Linear spin-wave theory}

\begin{figure}[b]
\includegraphics[width=3cm]{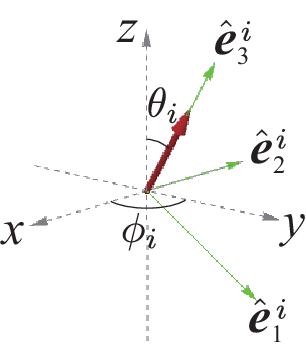}%
\caption{
Local orthogonal coordinate at the $i$th site spanned by unit vectors $\hat{\bm e}_i$ $(i=1, 2, 3)$. 
The unit vector $\hat{\bm e}^{i}_3$ is set along the magnetic-moment direction at each site. 
}
\label{fig:local_e123}
\end{figure}

To analyze the dynamical structure factor as well as the excitation energy in the hedgehog-antihedgehog order in the 1/1 AC, we employ the linear spin-wave theory. 
Since the hedgehog state and antihedgehog state are the non-collinear as well as non-coplanar magnetic state, 
it is convenient to introduce the orthogonal coordinate at each Tb site.
As shown in Fig.~\ref{fig:local_e123},  
the local coordinate at the $i$th Tb site is spanned by the unit vectors $\hat{\bm e}^{i}_{\beta}$~$(\beta=1,2$, and $3)$,  where $\hat{\bm e}^{i}_3$ points to the magnetic-moment direction. 
The unit vector 
$\hat{\bm e}^{i}_{\beta}$ and the unit vector $\hat{\bm r}_{\alpha}$ in the global coordinate $(\hat{\bm r}_1\equiv\hat{\bm x}, \hat{\bm r}_2\equiv\hat{\bm y}$, and $\hat{\bm r}_3\equiv\hat{\bm z})$  are related as 
\begin{eqnarray}
\hat{\bm r}_{\alpha}=R_{\alpha\beta}^{i}\hat{\bm e}_{\beta}^{i}, 
\label{eq:r_e}
\end{eqnarray}
where the direction of $\hat{\bm e}^{i}_3$ is denoted by the polar angles $(\theta_i, \phi_i)$ and $R^{i}$ is the rotation matrix defined by~\cite{Haraldsen}, 
\begin{eqnarray}
R^{i}=
\begin{bmatrix}
\cos\theta_i\cos\phi_i & -\sin\phi_i & \sin\theta_i\cos\phi_i \\
\cos\theta_i\sin\phi_i & \cos\phi_i & \sin\theta_i\sin\phi_i \\
-\sin\theta_i & 0 & \cos\theta_i
\end{bmatrix}. 
\label{eq:Rmat}
\end{eqnarray}
Then, the ``spin" interaction between the $i$th and $j$th sites in the first term of Eq.~(\ref{eq:H}) is written as 
\begin{eqnarray}
\sum_{\langle i,j\rangle}J_{i,j}({\bm S}_i\cdot{\bm e}_{\alpha}^i)({\bm S}_j\cdot{\bm e}_{\beta}^j)\sum_{\gamma}R_{\alpha,\gamma}^{i}R_{\gamma,\beta}^{j}. 
\label{eq:hamil2}
\end{eqnarray}
After substituting the relations    
${\bm S}_i\cdot\hat{\bm e}_1^i=(S_i^{+}+S_i^{-})/2$ and ${\bm S}_i\cdot\hat{\bm e}_2^i=(S_i^{+}-S_i^{-})/(2i)$ into Eq.~(\ref{eq:hamil2}), where $S_i^{+}$ is the  raising ``spin" operator and $S_i^{-}$ is the lowering ``spin" operator, the Holstein-Primakoff transformation~\cite{HP} is applied to $H$. 
In this way, the ``spin'' operators in Eq.~(\ref{eq:H}) are transformed into the boson operators as 
$S_i^{+}=\sqrt{2S-n_i}a_i$, $S_i^{-}=a_i^{\dagger}\sqrt{2S-n_i}$ and ${\bm S}_i\cdot\hat{\bm e}_3^i=S-n_i$ with $n_i\equiv a_i^{\dagger}a_i$
where $a^{\dagger}_i$~$(a_i)$ is the creation (annihilation) operator of the boson at the $i$th site. 
We keep the quadratic terms with respect to $a_i^{\dagger}$ and $a_i$, which are speculated to be valid at least for the ground state. 
Since the hedgehog state and antihedgehog state have non-collinear ``spin'' alignments, the anomalous terms such as $a_{i}a_{j}$ and $a^{\dagger}_{i}a^{\dagger}_{j}$ emerge 
besides the normal terms  $a^{\dagger}_{i}a_{j}$ and $a_{i}a^{\dagger}_{j}$. 

As noted below Eq.~(\ref{eq:Fs}), 
the position vector of the $i$th site is expressed as ${\bm r}_i={\bm R}_j+{\bm r}_{0m}$ where ${\bm R}_j$ denotes the position of the center of the $j$th unit cell and ${\bm r}_{0m}$ $(m=1, 2, \cdots, 24)$ denotes the position of the $m$th Tb site inside the unit cell. Hence, by carrying out the Fourie transformations $a_{i}=a_{j,m}=\frac{1}{\sqrt{N_L}}\sum_{\bm q}e^{i{\bm q}\cdot({\bm R}_j+{\bm r}_{0m})}a_{{\bm q},m}$ and $a^{\dagger}_{i}=a^{\dagger}_{j,m}=\frac{1}{\sqrt{N_L}}\sum_{\bm q}e^{-i{\bm q}\cdot({\bm R}_j+{\bm r}_{0m})}a^{\dagger}_{{\bm q},m}$  with ${\bm q}$ being the wave vector, the spin-wave Hamiltonian is expressed as the boson operators $a^{\dagger}_{{\bm q},m}$ and $a_{{\bm q},m}$. 
Then, by executing the para-unitary transformation~\cite{Colpa} i.e., the Bogoliubov transformation, the spin-wave Hamiltonian is diagonalized as 
\begin{eqnarray}
H=\sum_{\bm q}\sum_{m'=1}^{24}\omega_{m'}({\bm q})b^{\dagger}_{{\bm q},m'}b_{{\bm q},m'}, 
\label{eq:H_SW}
\end{eqnarray}
where $\omega_{m'}({\bm q})(>0)$ is the energy of the $m'$th spin wave. Here, 
$b^{\dagger}_{{\bm q},m'}$~$(b_{{\bm q},m'})$ is the creation (annihilation) operator of the boson which is composed of the linear combination of $a^{\dagger}_{{\bm q},m}$ and $a_{{\bm q},m}$.

\subsection{Energy dispersion of excitation from hedgehog-anti-hedgehog order}

\begin{figure}[t]
\includegraphics[width=7cm]{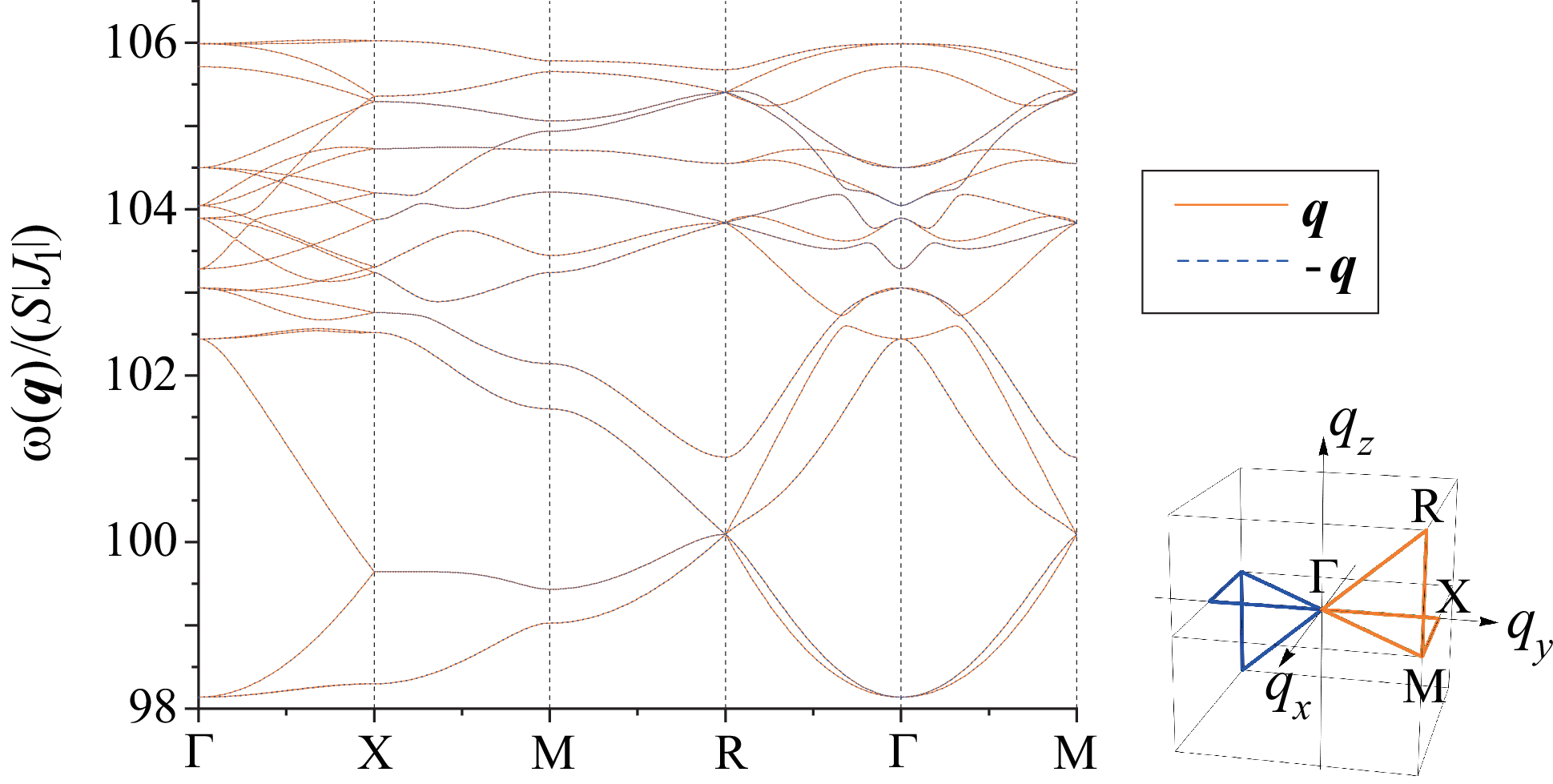}%
\caption{
The energy dispersion of the excitation in the hedgehog-antihedgehog order in the 1/1 AC 
for $D=50$  
at $J_1=-1.0$ and $J_2=-0.1$.  
The inset 
illustrates the Brillouin zone of the cubic unit cell with the side length $\frac{2\pi}{a}$. 
}
\label{fig:w_q}
\end{figure}

We analyze the energy dispersion of the excitation $\omega_{m'}({\bm q})$ in the hedgehog-antihedgehog order in the 1/1 AC. 
We have performed the numerical calculations for $J_1=-1.0$ and $J_2=-0.1$ in the systems for $N_L=N_1^3$ with $N_1=8$, 16, 32, 64, and 128. 
The results of the excitation energy $\omega_{m'}({\bm q})$ for $N_1=64$ and $128$ are seen as the same and we show 
the result in the system with $N_1=128$ for 
for $D=50$ in Fig.~\ref{fig:w_q}.
Here ${\bm q}$ is plotted along the symmetry line (orange line) in the Brillouin zone 
illustrated as the frame box with the side length $\frac{2\pi}{a}$ in the inset.  
The coordinate of each symmetry point in the reciprocal space is as follows: $\Gamma$: $(0, 0, 0)$, X: $\frac{2\pi}{a}(\frac{1}{2}, 0, 0)$, M: $\frac{2\pi}{a}(\frac{1}{2}, \frac{1}{2}, 0)$, and R:  $\frac{2\pi}{a}(\frac{1}{2}, \frac{1}{2},\frac{1}{2})$.  
Since there exist 24 Tb sites in the unit cell [see Fig.~\ref{fig:HAH}(c)], the 24 energy bands for the excitation appear in Fig.~\ref{fig:w_q}. 
Because of the uniaxial anisotropy $D$ arising from the CEF, the lowest excitation energy appears 
at the finite energy. 
{
It is noted that the energy gap opens for $D\ge 1.7836$ which was obtained by the linear-spin-wave calculation assuming the hedgehog-antihedgehog order as the ground state.}
{
It is cofirmed that the overall features of the energy dispersions of the magnetic excitations $\omega({\bm q})$ shown in Fig.~\ref{fig:w_q} are the same as long as $D\gsim 5$.
}



In Fig.~\ref{fig:w_q}, 
we also plot $\omega_{m'}(-{\bm q})$ as the blue dashed line for ${\bm q}$ along the symmetry line illustrated in the inset, where $-{\bm q}$ follows the blue line in the inset. 
We see that all the blue dashed lines coincide with the orange solid lines, which indicates that the reciprocal energy dispersion $\omega_{m'}({\bm q})=\omega_{m'}(-{\bm q})$ is realized.
This is in sharp contrast to the nonreciprocal energy dispersion $\omega_{m'}({\bm q})\ne\omega_{m'}(-{\bm q})$ shown in the uniform hedgehog order in the QC~\cite{WHH2022} and 1/1 AC~\cite{W2023}.  
This can be intuitively understood in terms of the symmetry operation as follows.

\begin{figure}[t]
\includegraphics[width=6cm]{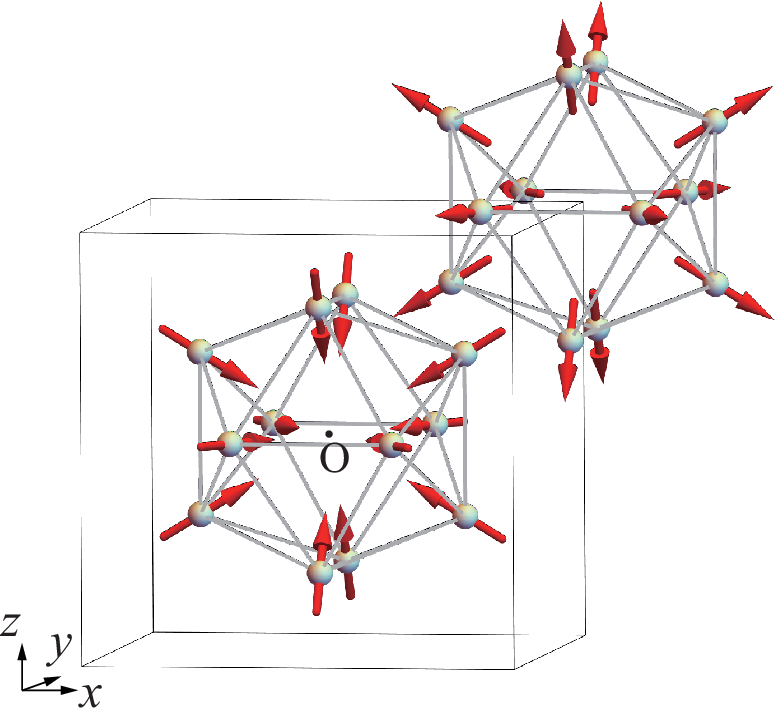}%
\caption{
In the 1/1 AC, 
the antihedgehog state on the center IC and the hedgehog state on the corner IC in the bcc unit cell illustrated by the frame box.
}
\label{fig:AHH}
\end{figure}

Let us consider the global coordinate in Fig.~\ref{fig:HAH}(c) whose origin is set to be the center of the unit cell. 
When the magnetic moment located at the position ${\bm r}$ in Fig.~\ref{fig:HAH}(c) is spatially inverted as $-{\bm r}$ with the moment direction being kept, the hedgehog state on the central IC is transformed to the antihedgehog state and the antihedgehog state on the corner IC is transformed to the hedgehog state, as shown in Fig.~\ref{fig:AHH}. 
Since the magnetic structure shown in Fig.~\ref{fig:AHH} is regarded to be equivalent to that shown in Fig.~\ref{fig:HAH}(c) as the global structure of the whole crystal, the spatial inversion symmetry is not broken by the hedgehog-antihedgehog ordering.  
Hence, the reciprocal energy dispersion is understood to appear in Fig.~\ref{fig:w_q}.

On the contrary, in the case of the uniform hedgehog order in the 1/1 AC, when the magnetic moment located at the position ${\bm r}$ is spatially inverted as $-{\bm r}$ with the moment direction being kept, 
both the hedgehog states on the central IC and the corner IC are transformed to the antihedgehog states. Since the uniform hedgehog state and the uniform antihedgehog state are different magnetic states each other, this implies that the spatially inversion symmetry is broken by the uniform hedgehog ordering. 
The same is applied to the uniform hedgehog order in the QC discussed in Ref.~\cite{WHH2022}. 
Hence, nonreciprocal energy dispersions emerge, i.e., $\omega_{m'}({\bm q})\ne\omega_{m'}(-{\bm q})$, in these cases.   

From these analyzes, it turns out that the pair formation of the hedgehog state and the antihedgehog state in the 1/1 AC [see Fig.~\ref{fig:HAH}(c) and Fig.~\ref{fig:AHH}] preserves the inversion symmetry. 

\subsection{Dynamical structure factor of magnetism}

\begin{figure}[t]
\includegraphics[width=7cm]{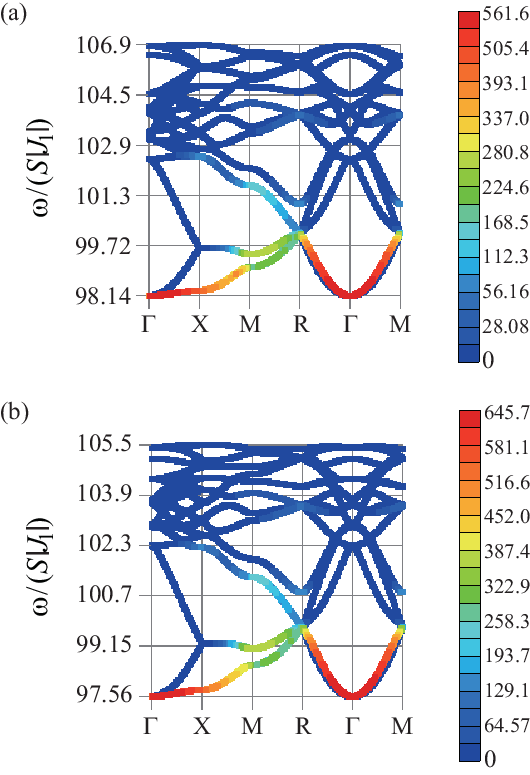}%
\caption{
The intensity plot of 
the dynamical structure factor $S_{\perp}({\bm q},\omega)$ 
for $D=50$ 
at $J_1=-1.0$ and (a)$J_2=-0.1$ and (b) $J_2=-0.2$ 
where ${\bm q}$ is plotted along the symmetry line illustrated in the inset of Fig.~\ref{fig:w_q}. 
}
\label{fig:Sperp}
\end{figure}

On the basis of the linear spin-wave theory, 
we analyze the dynamics of the magnetic excitation in the hedgehog-antihedgehog order in the 1/1 AC. 
The magnetic dynamical structure factor is defined by
\begin{eqnarray}
S_{\alpha\beta}({\bm q}, \omega)=-\frac{1}{\pi}{\rm Im}\langle{\rm GS}|S_{{\bm q}\alpha}\frac{1}{\omega+E_{0}-H+i\eta}S_{-{\bm q}\beta}|{\rm GS}\rangle, 
\label{eq:Gij}
\end{eqnarray}
where $|{\rm GS}\rangle$ is the ground state with the energy $E_{0}$ and $S_{{\bm q}\alpha}$ is the Fourier transformation of the ``spin'' operator defined by 
$S_{{\bm q}\alpha}=\frac{1}{\sqrt{N}}\sum_{i=1}^{N}e^{-i{\bm q}\cdot{\bm r}_i}S_{i\alpha}$. 
In this study, we set $\eta=10^{-3}$. 
Experimentally, the magnetic dynamical structure factor can be observed by the neutron measurement.
The intensity of inelastic neutron scattering is expressed by the dynamical structure factor 
\begin{eqnarray}
S_{\perp}({\bm q},\omega)=\sum_{\alpha,\beta=x,y,z}(\delta_{\alpha\beta}-\hat{q}_{\alpha}\hat{q}_{\beta})S_{\alpha\beta}({\bm q},\omega), 
\label{eq:S_perp}
\end{eqnarray}
where ${\bm q}$ is the incident wave vector and  $\hat{q}_{\alpha}$ is defined as $\hat{q}_{\alpha}\equiv q_{\alpha}/|{\bm q}|$~\cite{Squires}. 
We have performed the numerical calculations of $S_{\perp}({\bm q},\omega)$ for $J_1=-1.0$ and $J_2=-0.1$ in the systems for $N_L=N_1^3$ with $N_1=8$, 16, 32, 64, and 128. 
We have confirmed that the results in $N_1=64$ and 128 are almost the same and 
the result for $D=50$
in the system with $N_1=128$ is shown in 
Fig.~\ref{fig:Sperp}.
Here, the intensity of  $S_{\perp}({\bm q},\omega)$ is plotted   
for ${\bm q}$ along the  symmetry line (orange line) illustrated in the inset of Fig.~\ref{fig:w_q}. 


In Fig.~\ref{fig:Sperp}, 
the high intensity appears in the lowest-energy branch along the $\Gamma$-X line and also in the 
second-lowest  
energy branch along the R-$\Gamma$-M line. The relatively-high to moderate intensities appear in the lowest- and second-lowest-energy branches along the X-M-R line. 
{
We confirmed that this feature appears for $D\gsim 2$ by calculating $S_{\perp}({\bm q},\omega)$ on the basis of the linear-spin-wave theory applied to the hedgehog-antihedgehog order.
}

Since it was shown that the hedgehog-antihedgehog order is realized for $J_2/J_1< 0.284$ in the effective magnetic model which corresponds to the large-$D$ limit of the model (\ref{eq:H}) in Ref.~\cite{WSR}, we have discussed the results of the linear spin-wave theory for $J_1=-1.0$ and $J_2=-0.1$ as the typical parameters. 
Here, the $J_2$ dependence of the results is discussed. 
We have calculated the dynamical structure factor $S_{\perp}({\bm q},\omega)$ as well as the excitation energy $\omega_{m'}({\bm q})$ for various $J_2/J_1< 0.284$ with the FM interaction $J_1<0$.

{
We calculated $S_{\perp}({\bm q},\omega)$ of the excitation from the hedgehog-antihedgehog order for $J_2=-0.2$, as shown in Fig.~\ref{fig:Sperp}(b). The energy dispersion and the intensity of $S_{\perp}({\bm q},\omega)$ are almost similar to the result for $J_2=-0.1$ shown in Fig.~\ref{fig:Sperp}, where all blanches are slightly shifted to smaller $\omega$, i.e., the lowest-excitation energy is $\omega/(S|J_1|)=97.56$ and the highest-excitation energy is $\omega/(S|J_1|)=105.5$. 
}

\section{Summary and Discussion}

We have theoretically analyzed the dynamical structure factor as well as the static structure factor in the hedgehog-antihedgehog order in the 1/1 AC. 

We have shown that 
the static structure factor $F_s({\bm q})$ is generally expressed as the convolution form  $F_s({\bm q})=F_L({\bm q})S_{2{\rm IC}}({\bm q})$ of the lattice structure factor $F_L({\bm q})$ and the magnetic structure factor of the 2 ICs $S_{2{\rm IC}}({\bm q})$ in the unit cell. 
In the bulk limit, $F_{s}({\bm q})$ can have a finite value for ${\bm q}=\frac{2\pi}{a}(h, k, l)$ with integer $h$, $k$, and $l$ because of the requirement from the lattice structure factor $F_L({\bm q})$.  
By the numerical calculation for the hedgehog-antihedgehog ordered state, we have shown that the systematic absence of reflection $F_s({\bm q})=0$ occurs for ${\bm q}=\frac{2\pi}{a}(h,k,l)$ 
{
for case that all $h$, $k$, and $l$ are even integers or one of $h$, $k$, and $l$ is even integer}. 
This extinction rule and the distribution of the intensities of $F_s({\bm q})$ revealed in this study is useful for identifying the hedgehog-antihedgehog order experimentally. 

On the basis of the linear spin-wave theory, we have analyzed the excitation energy and the dynamical structure factor 
{
by introducing the effective magnetic model with uniaxial anisotropy $D$ arising from the CEF.
}
The results 
{
for large $D$ 
}
show that the reciprocal energy dispersion emerges, which is understood as the symmetry argument that the hedgehog-antihedgehog ordering preserves the spatial inversion symmetry. 
This is in sharp contrast to the emergence of the non-reciprocal excitations in the uniform hedgehog order in the 1/1 AC and QC where the spatial inversion symmetry is broken by the hedgehog ordering. 
%
The high intensity of $S_{\perp}({\bm q},\omega)$ appears in the low-energy branch in the vicinity of the $\Gamma$ point along the $\Gamma$-X line and the R-$\Gamma$-X line.

So far, the hedgehog and antihedgehog states have been theoretically shown in the 1/1 AC and QC~\cite{WSR,WPNAS} but have not been observed experimentally. 
To observe the dynamical as well as the static structure factor shown in this paper, it is necessary to identify the material which shows the hedgehog-anti-hedgehog state in the 1/1 AC. 

The CEF in the Au-SM-Tb (SM=Si, Ge, Al, and Ga) was analyzed theoretically on the basis of the point charge model in refs.~\cite{WSR}  and \cite{WPNAS}. It was shown that as the ratio of the valences of the screened ligand ions $\alpha\equiv Z_{\rm SM}/Z_{\rm Au}$ surrounding the Tb$^{3+}$ ion varies, the direction of the magnetic easy axis at each Tb site changes. This suggests that by changing the compositions of the non-rare earth elements in the ternary compounds Au-SM-Tb, there is the possibility that the hedgehog state and anti-hedgehog state are realized. It is also noted that the CEF in the Au-Al-Yb was analyzed theoretically in ref.~\cite{WK2021}. It was shown that the tendency of the $\alpha$ dependence of the magnetic easy axis in the CEF ground state is different from that in the Au-SM-Tb systems. Hence, there also exists the possibility that in the 1/1 AC Au-SM-R with the other rare-earth element(s), the hedgehog-anti-hedgehog state is realized. Hence, the synthesis of the rare-earth based Au-SM-R compounds is interesting future subject to realize the hedgehog-anti-hedgehog state.

Since the effective model for magnetism (\ref{eq:H}) is considered to be relevant to the broad range of the rare-earth based ACs not only for Tb but also for the other rare-earth atoms, the results of the dynamical as well as static structure factor clarified in the present study are expected to be useful for identifying the hedgehog-antihedgehog order in the 1/1 AC. 
Moreover, the observation of the magnetic dynamics of the ``spin'' structures on the IC has neither been reported in the icosahedral QCs and ACs, thus far.  
The present study is expected to stimulate future experiments to detect the dynamical as well as static structure factor in the unique magnetic structure with the non-collinear and non-coplanar alignments of ``spins'' in the icosahedral 1/1 ACs and also QCs. 

\begin{acknowledgments}

The author thanks H. Harima and T. Ishimasa for valuable discussion. 
The author greatly acknowledges K. Momma for providing the VESTA to draw magnetic moments in the 1/1 AC with kind instructions.
This work was supported by JSPS KAKENHI Grant Numbers JP18K03542, JP19H00648, JP22H04597, JP22H01170, and JP23K17672.
\end{acknowledgments}

\end{document}